\pdfoutput=1
\RequirePackage{ifpdf}

\documentclass{JINST}
\usepackage{cite}

\title{The Effects of Dissolved Methane upon Liquid Argon Scintillation Light}

\author{B.J.P. Jones$^a$\thanks{Corresponding Author}, T. Alexander$^b$, H.O. Back$^c$, G. Collin$^a$, J.M. Conrad$^a$, A. Greene$^a$, T. Katori$^a$, S. Pordes$^d$, M. Toups$^a$.\\
\llap{$^a$}Massachusetts Institute of Technology,
  77 Massachusetts Avenue, Cambridge, MA 02139, USA\\
  E-mail: \email{bjpjones@mit.edu}\\
\llap{$^b$}University of Massachusetts at Amherst,
  710 N Pleasant St. Amherst MA 01003, USA  \\
\llap{$^c$}Princeton University,
   Princeton, NJ 08540, USA \\ 
\llap{$^d$}Fermi National Accelerator Laboratory,
  Batavia, IL 60510, USA}

\abstract{In this paper we report on measurements of the effects of dissolved methane upon argon scintillation light.  We monitor the light yield from an alpha source held 20~cm from a cryogenic photomultiplier tube (PMT) assembly as methane is injected into a high-purity liquid argon volume. We observe significant suppression of the scintillation light yield by dissolved methane at the 10 part per billion (ppb) level.  By examining the late scintillation light time constant, we determine that this loss is caused by an absorption process and also see some evidence of methane-induced scintillation quenching at higher concentrations (50-100 ppb).  Using a second PMT assembly we look for visible re-emission features from the dissolved methane which have been reported in gas-phase argon methane mixtures, and we find no evidence of visible re-emission from liquid-phase argon methane mixtures at concentrations between 10 ppb and 0.1\%.
}


\begin{document}

\section{Motivations for Studying Methane in Liquid Argon}
Many current- and future-generation neutrino and dark matter experiments use liquid argon as an active medium \cite{Akiri:2011dv, Rubbia:2012jr, Jones:2011ci, Menegolli:2012jq, Boulay:2012hq, Rielage:2012zz, Akimov:2012nr}.  The detection of weakly interacting particles-of-interest in such a detector involves collecting ionization charge and/or scintillation light produced in the argon bulk.  In order to achieve a large target mass with a reasonable number of sensitive detector elements, the detection of both light and charge typically takes place some distance from the interaction point.  For charge collection, this is achieved by drifting free electrons in an electric field, as in a time projection chamber (TPC).  For light detection, sensitive elements such as PMTs \cite{Briese:2013wua} or light guides \cite{Baptista:2012bf} are placed around the boundaries of an argon target volume.  Because argon is transparent to its own scintillation light, some fraction of the isotropically produced scintillation light is captured by these sensitive elements. 

Both charge and light production and propagation are known to be affected adversely by small concentrations of dissolved impurities. The commercially available argon which is used in most liquid argon detectors is manufactured by the distillation of air, and typically arrives from the vendor with part-per-million (ppm) levels of contaminants such as water, oxygen and nitrogen \cite{AirGasSpec}.  For a TPC detector, water and oxygen must be removed in a subsequent filtering process to the hundred parts per trillion level in order not to damage the free electron lifetime \cite{Baibussinov:2009gs}.  In a scintillation detector, not only water and oxygen, but also nitrogen must be controlled at the ppm level in order to prevent scintillation quenching \cite{Acciarri:2008kv,Acciarri:2008kx,Acciarri:2009xj} and absorption \cite{Jones:2013xy}. 

A new generation of dark matter experiments, which require highly radiopure argon \cite{Akimov:2012nr}, plan to use argon that is not distilled from air, but rather Underground Argon (UAr) extracted from underground sources.  This argon has the advantage of having low levels of the radioactive isotope $^{39}Ar$ \cite{Xu:2012xe}, a source of background in such detectors.  The UAr is extracted from carbon dioxide wells that are comprised of 96\% carbon dioxide and 2\% nitrogen, alongside sub-percent levels of other gasses, including 5,700 ppm methane and 600 ppm argon.  Locally at the CO$_2$ well, an Ar-He-N$_2$ mixture is extracted from the crude CO$_2$ gas, with methane levels below the sensitivity on the \textit{in situ} gas analyzer \cite{Back:2012rw}. The gas mixture is sent to the Fermi National Accelerator Laboratory for further purification, where, after He removal, cryogenic distillation removes the N$_2$ and purified UAr is produced \cite{Back:2012la}. This cryogenic distillation process not only concentrates the UAr, but also has the effect of concentrating any residual trace amounts of methane. It is therefore critical to know what concentration of methane is permissible in the UAr scintillation experiments in order to understand what purification techniques are required to remove the methane, which can be technically difficult.

On the other hand, mixed methane and liquid argon TPCs have been proposed as detectors for low-energy neutrino physics \cite{Aprile1987273}.  The benefit provided by the methane admixture is that free protons are present in the detector as interaction targets.  This reduces the threshold for the interactions of low-energy neutrinos, such as those from the sun, or from supernovae, or man-made neutrinos from decay-at-rest sources.  Free protons may also make neutron captures detectable in a liquid argon scintillation detector, allowing for coincidence tagging of low-energy antineutrino inverse-beta-decay interactions ($\overline{\nu}_e + p \rightarrow e^+ + n$).

The effects of methane upon charge collection have been shown to be minimal, even at few-percent level concentrations \cite{Aprile1987273}.  However, the effects of methane upon liquid argon scintillation light have not been investigated previously.  As such, no specification on the allowed methane concentration for a scintillation detector currently exists.  World data on the VUV absorption cross section of pure methane, taken at room temperature and pressure, suggest that significant absorption may be expected at relatively low concentration levels \cite{WorldUVData}.  Studies of scintillation light from gaseous argon-methane mixtures, however, have reported a visible emission feature at a wavelength of a 431 nm \cite{Siegmund:1981un, 4336365}.  This motivates a determination of the effects of methane impurities in liquid argon detectors, operating under realistic scintillation detection conditions. 

In this paper we examine the effects of methane upon the light yield from alpha particles detected in a liquid argon scintillation detector, and quantify the concentration scale to which methane must be removed in order to prevent significant scintillation light losses relative to pure argon.

\section{Experimental Configuration for this Study}

\begin{figure}[tb]
\begin{centering}
\includegraphics[width=1.0\columnwidth]{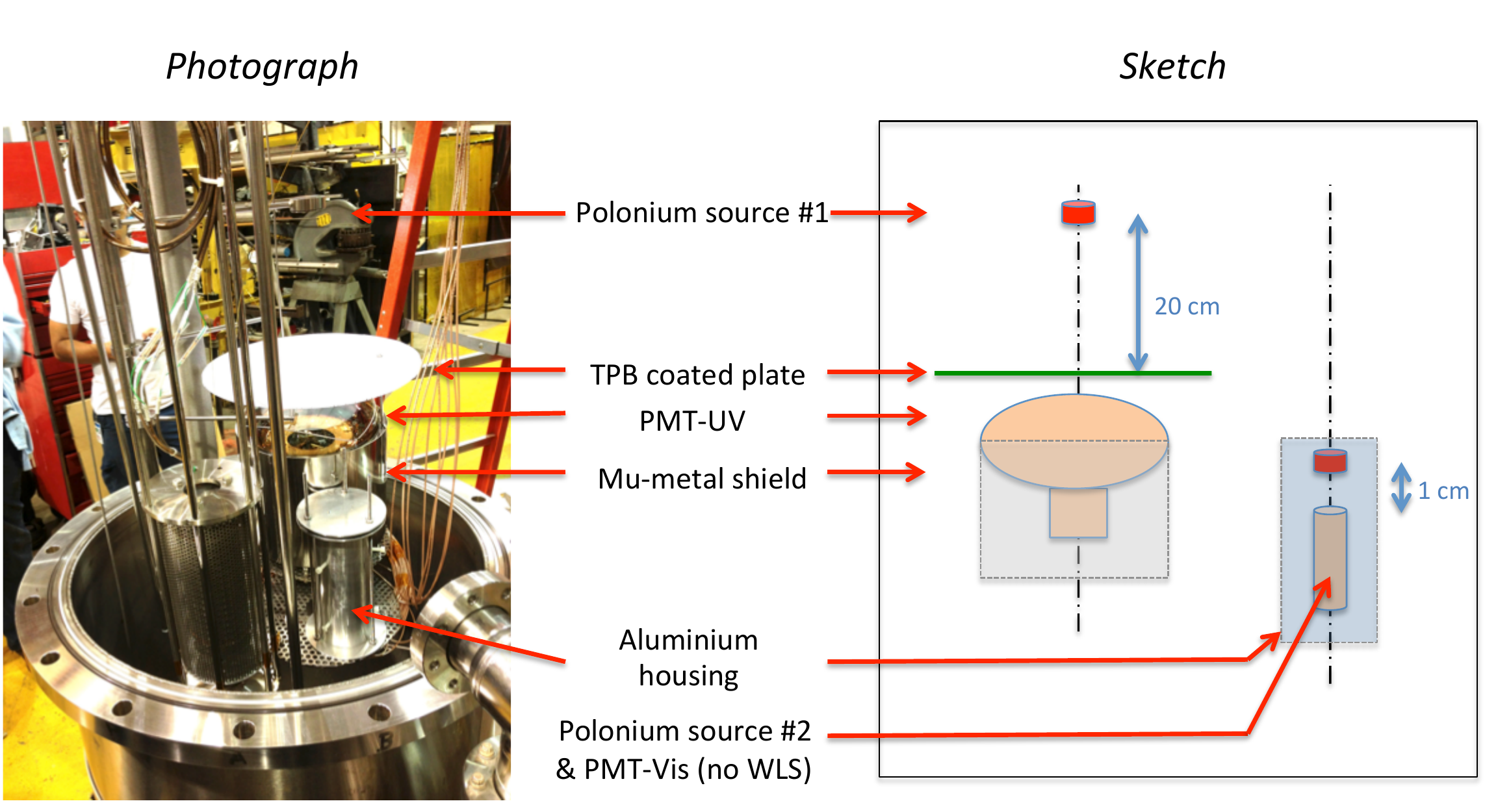}
\par\end{centering}

\caption{A labeled photograph and sketch showing the internal Bo configuration used in this study.  \label{fig:PhotographOfAparatus}}
\end{figure}

These tests were performed using the Bo cryostat and gas injection system, which was previousely used to study the effects of nitrogen contamination in argon in \cite{Jones:2013xy}.  The cryostat is a vacuum insulated cylinder with a diameter of 55.9 cm which can be filled to a level of approximately 100 cm with high purity argon, supplied via a system of molecular seives and regenerable filters \cite{Curioni:2009rt}.  For this study, the cryostat was filled with argon to a level of 76$\pm$1 cm and maintained at a pressure of 12 $\pm$ 0.2 psig by a liquid nitrogen condenser system.

Inside the cryostat are situated two cryogenic PMT assemblies.  The first assembly, which will be referred to hereafter as PMT-UV, is comprised of a wavelength shifting plate supported above an 8-inch cryogenic Hamamatsu R5912-02mod PMT \cite{Briese:2013wua}, which is mounted inside a mu-metal shield.  The wavelength shifting plate is a 12-inch disc of clear acrylic upon one side of which is a coating of tetraphenyl butadiene (TPB) suspended in a polystyrene matrix \cite{Jones:2012hm, Gehman:2011xm, Gehman:2013lsx}.  The purpose of the plate is to capture the 128 nm argon scintillation light and re-emit at visible wavelengths, with an emission spectrum peaked around 450 nm where PMT quantum efficiency is high \cite{HamamatsuPMT}.  A polonium disc alpha source is held 20~cm from the face of PMT-UV, providing a source of scintillation light in the form of monoenergetic 5.3 MeV alpha particles.  Monitoring the light yield at PMT-UV as methane is added indicates to what extent the 128~nm argon scintillation light has been absorbed or quenched by the injected methane. In addition to scintillation light from the polonium alpha source, PMT-UV also detects the scinitllation light produced by cosmic rays traversing the Bo volume, and would also be sensitive to visible re-emission features if any were present.

The second assembly, which will be referred to hereafter as PMT-Vis, is comprised of a 2-inch Hamamatsu R7725-mod PMT, housed inside an aluminium cylinder.  The 2-inch PMT has no accompanying wavelength shifter and is shielded by the aluminium cylinder from visible light generated by the PMT-UV wavelength shifting plate.  A second polonium disc alpha source is held 1 cm from the face of the PMT-Vis.  Openings in the base and around the top edge of the aluminium cylinder provide space for argon to flow through the two inch assembly to ensure that there is mixing throughout the system.  Since there is no wavelength shifter, this PMT is insensitive to the 128 nm argon scintillation light, so can be used to test for the emergence of possible visible re-emission features.  The 5.3MeV alpha particles produced by the polonium source have a range of only 50 $\mathrm{\mu m}$ in liquid argon, so can be treated as point light sources for our purposes.  With a source to PMT distance of 1 cm, the 2-inch diameter PMT face subtends a solid angle of more than 30\% to these point light sources. The PMT quantum efficiency is around 20\% at visible wavelengths.  Accounting for these two factors we expect to be able to collect at least 6\% of emitted visible photons.  Therefore an emission feature as small as 10 visible photons / MeV, which produces an average of 53 visible photons per alpha particle would lead to a detected distribution with a Poisson mean of three photoelectrons.  This signal would be clearly visible above background in our apparatus.

Signal and high voltage for each PMT are carried by common RG-316 type coaxial cables, and split outside of the cryostat by a HV splitter unit.  The signal is connected to a Tektronix DPO5000 oscilloscope terminated at 50$\Omega$ and sampling at 1 gigasample per second. 

\begin{figure}[tb]
\begin{centering}
\includegraphics[width=1\columnwidth]{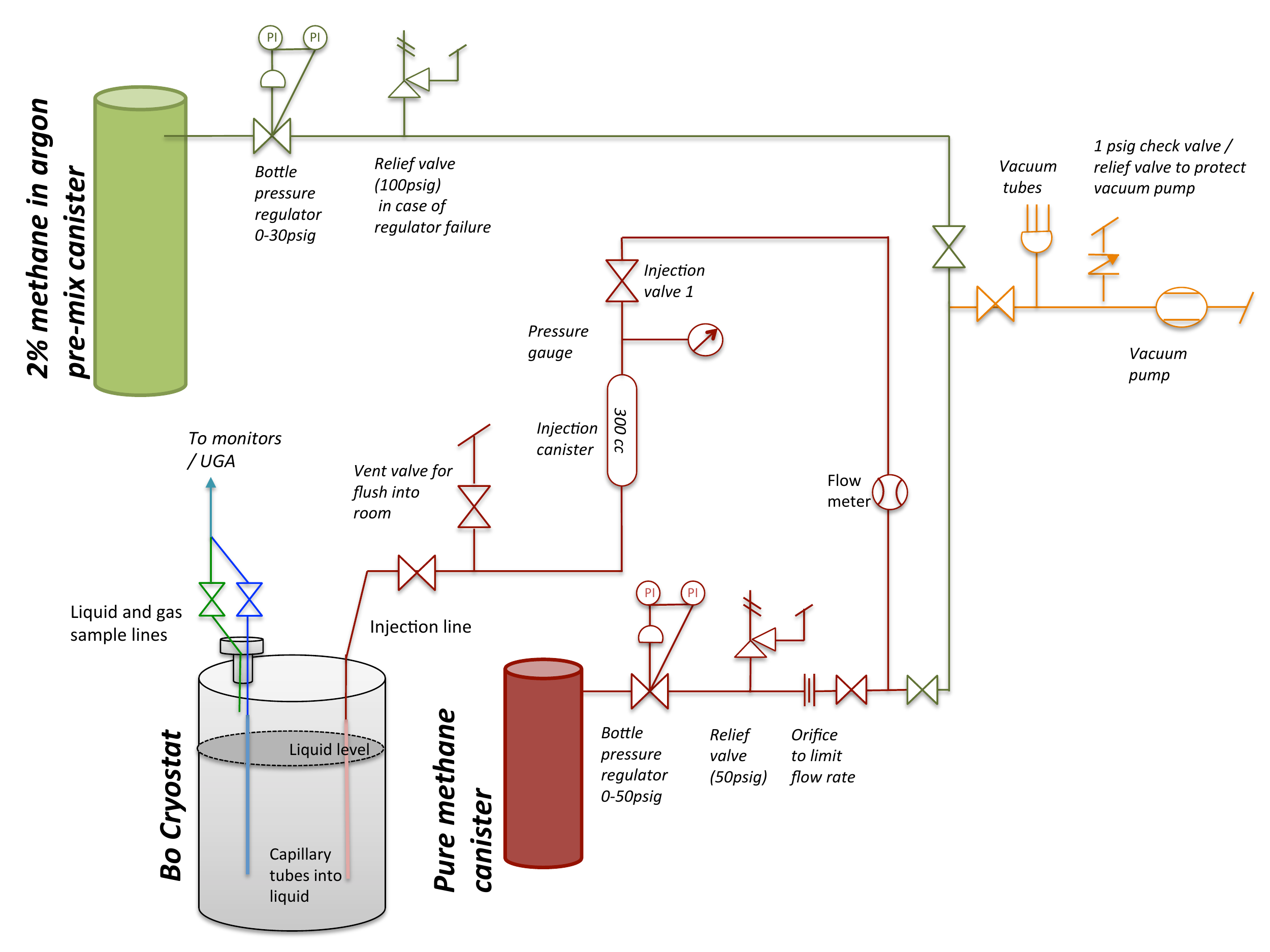}
\par\end{centering}

\caption{A diagram showing the layout of the sample and injection lines connected to the Bo cryostat for this study. The majority of the data reported in this paper were acquired by injecting from the 2\% methane in argon canister.  \label{fig:NitInjMethane}}
\end{figure}

Our measurement involves making injections of small amounts of methane gas into the liquid argon volume, and studying the light yield at both PMT-UV and PMT-Vis.  Methane injections are made via a 300 $\mathrm{cm^3}$ injection canister which can be charged to a known pressure with an injection gas, and then released through a capillary pipe into the liquid volume.  The injection pressure is measured using an analog pressure gauge which has a precision of 1 psig.  The internal Bo pressure into which the gas is released is recorded by an electronic pressure gauge with a precision of 0.1 psig. After being released into the Bo volume, the injected gas circulates and eventually reaches an equilibrium distribution between the liquid and vapor phases.  The apparatus also includes sample capillary lines for the liquid and vapor phases which can be used to feed monitors and gas analyzers when required.  A diagram showing the layout of the injection and sample lines used in this study is shown in figure \ref{fig:NitInjMethane}.

\begin{figure}[tb]
\begin{centering}
\includegraphics[width=1.0\columnwidth]{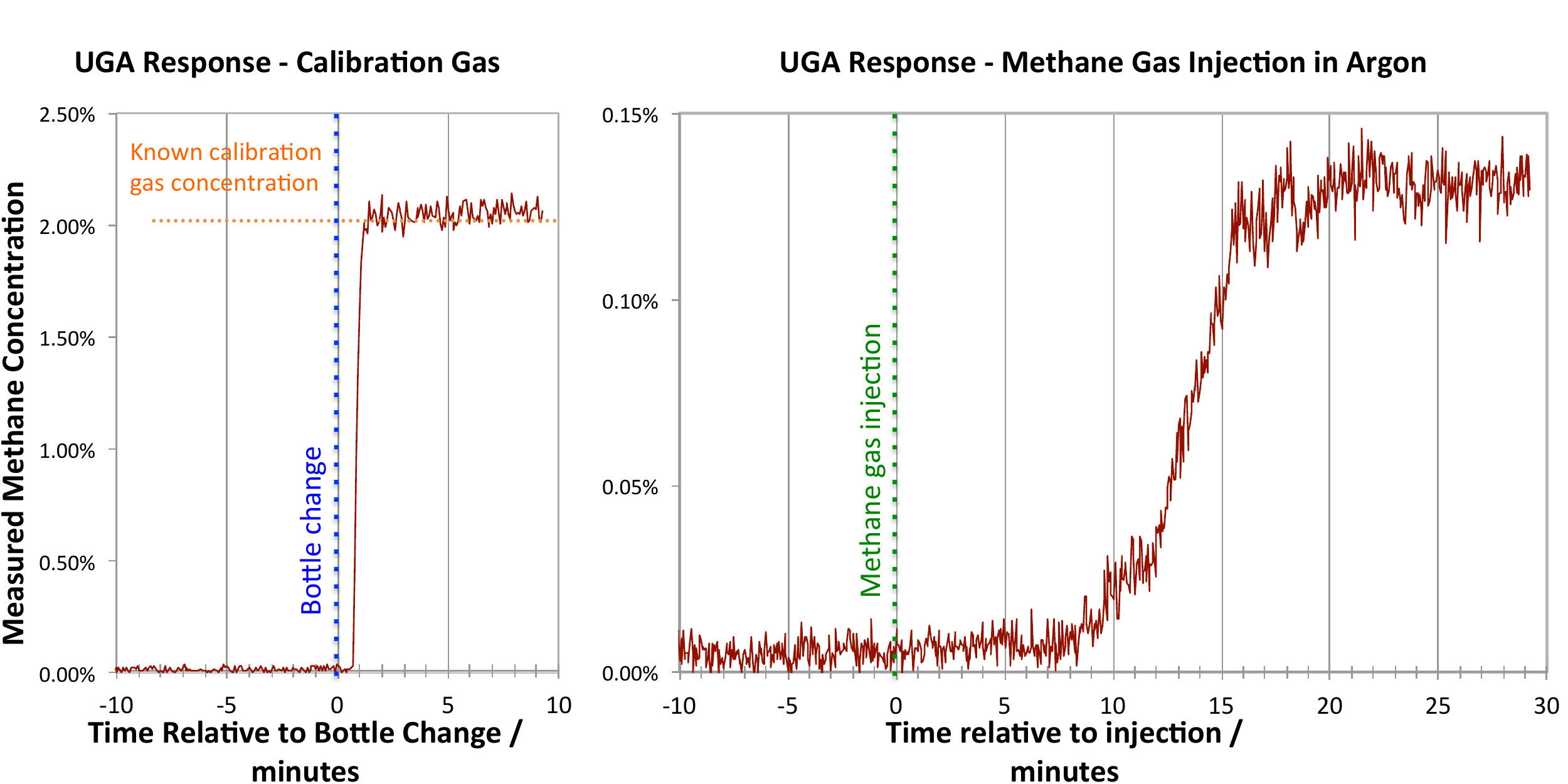}
\par\end{centering}

\caption{Left: Time taken for reported methane concentration by UGA to stabilize after swapping source from Bo sample line to calibration gas at the Bo cryostat location.  This quantifies the flow time in the sample piping and the lag time of the UGA.  Right: Time taken for reported methane concentration by UGA to stabilize after an injection of methane into pure argon in the Bo cryostat.  The equilibration time is approximately 20 minutes.\label{fig:EquilibrationPlots}}
\end{figure}

The equilibration time of methane in argon following an injection in this system determines the amount of time which must be left between methane injection and light measurement.  We measured the equilibration time of the system after a methane injection in a preliminary study using a universal gas analyzer (UGA) \cite{UGAReference}.  First, the gas analyzer was connected to the Bo liquid sample line with the cryostat held at 12~psig.  A stable methane concentration at the noise floor of the device, corresponding to around 0.01\% methane per mole\footnote{All concentrations reported hereafter will be implicitly given as molar ratios} was observed.  At a known time, the UGA piping was disconnected from the Bo cryostat and connected to a bottle of 2\% methane-in-argon calibration gas regulated at 12 psig.  The time taken for the methane concentration reported by the UGA to stabilize to 2\% was approximately one minute, which sets the timescale for flow of gas through the 30~m supply pipe and the lag time of the measuring device.  Then, with Bo reconnected and the methane reading stable at the noise floor once again, an injection of around 0.1\% methane was made into Bo.  The reported methane concentration by the UGA is seen to rise gradually, and then stabilize after approximately 20 minutes.  Since we already know that the flow time in the piping and the UGA response time is much shorter than this, we conclude that 20 minutes is the equilibration time for methane injected into the cryostat to reach an equilibrium distribution between the gas and the liquid phases.  These measurements can be seen in Figure \ref{fig:EquilibrationPlots}.  We allow a conservative 40 minutes of equilibration time between any injection and light yield measurement.  The concentrations of methane we investigate in this paper are far below the sensitivity of the UGA, and so we cannot use it to quantify the injected methane.  Our only concentration measurements come from the known volume and pressure of the injected gas, which is accompanied by relatively large systematic uncertainties.  We describe these below.

In order to achieve ppb level concentrations of methane in argon, we inject a pre-mixed gas of 2\% methane in argon via a 300 $\mathrm{cm}^3$ injection canister, charged to a known pressure with gas at room temperature and then discharged into the liquid.  To achieve injected methane concentrations at the 10~ppb level, the injection pressure must be set fairly low relative to the design pressure of the injection system.  The lowest methane concentration point requires a 15 psig injection into the 12 psig Bo liquid volume.  Using the ideal gas law, we can determine that the 3 psi of injected pre-mixed gas contains $\mathrm{5\times10^{-5}}$ moles of methane.  The Bo liquid volume is a cylinder of diameter 55.9 cm and height 76 cm, which corresponds to approximately 6530 moles of argon.  Thus a 15 psig injection into the 12 psig volume leads to an increase in methane concentration by 8 ppb. 

The dominant systematic uncertainties which affect the injected concentration are the injection pressure gauge precision of $\pm$ 0.5 psig, and the imprecisely known injection volume, which has an uncertainty of $\pm$ 15\% due to volume associated with valves and piping.  There are other sub-dominant contributions to the total uncertainty, which are enumerated in Table \ref{tab:SystUncertTable}.  Adding all the uncertainties in quadrature leads to an overall concentration uncertainty of 24\% for this lowest pressure injection.  The extent to which these uncertainties are correlated between injections is difficult to quantify in our apparatus, and as such we conservatively assign all concentrations the same worst-case fractional error of 24\%.  For future higher precision studies, the injection system may be improved by installing a smaller canister or higher precision pressure gauge to produce much more precisely known concentrations.  For the purposes of this paper, however, a logarithmic scan of the relevant concentration range is already very instructive.

\begin{table}[h]
\begin{centering}
\begin{tabular}{|c|c|c|c|}
\hline 
\textbf{Quantity} & \textbf{Value} & \textbf{Unit} & \textbf{Contribution to} \tabularnewline &&&\textbf{Concentration Uncertainty}\\
\hline
\hline
Injection pressure precision & $15\pm0.5$ &psig & $16.5\%$\tabularnewline
\hline 
Injection volume & $300\pm45$&$\mathrm{cm^3}$ & $15\%$\tabularnewline
\hline 
Gas temperature & $290\pm10$& K & $3.4\%$\tabularnewline
\hline 
Internal pressure stability & $12\pm0.2$&psig & $1.7\%$\tabularnewline
\hline 
Liquid level & $76\pm1$&cm & $1.3\%$\tabularnewline
\hline 
\hline
\textbf{Total injected} & $\mathbf{8.0\pm1.9}$ & \textbf{ppb} & $\textbf{23.7\%}$\tabularnewline
\hline

\end{tabular}
\par\end{centering}
\caption{Contributions to the systematic uncertainty of the injected methane concentration for a 15 psig injection into the 12 psig liquid volume.
 \label{tab:SystUncertTable}}
\end{table}

\section{The Effect of Methane upon Prompt Scintillation Light Yield \label{sec:PromptYieldSection}}

\begin{figure}[t]
\begin{centering}
\includegraphics[width=1.0\columnwidth]{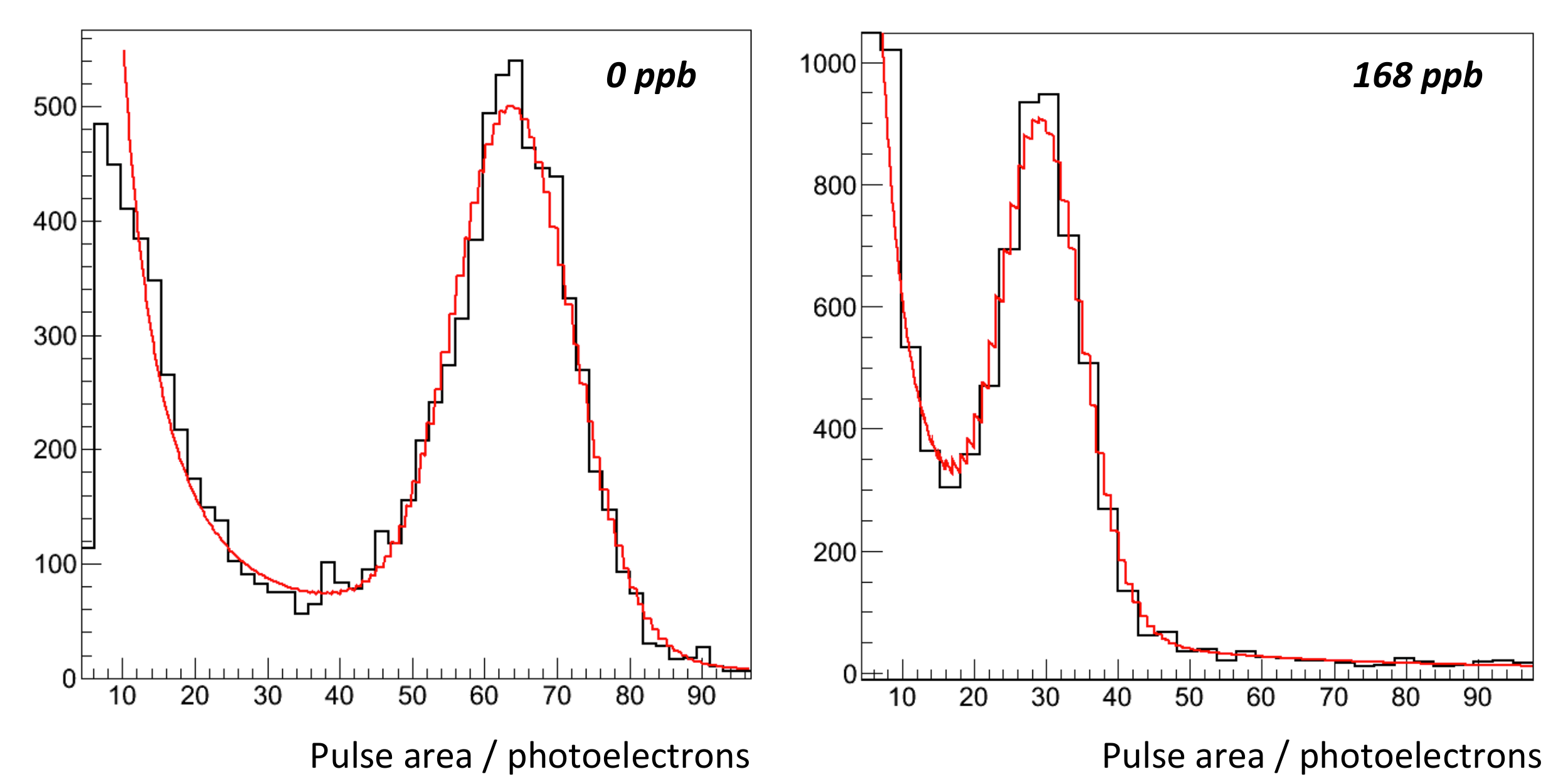}
\par\end{centering}

\caption{The PMT-UV pulse area spectrum and best fit function for the 0 ppb and 168 concentration points.  The fit function is a modified poissonian distribution superposed with a falling power law background, as described in \cite{Jones:2013xy}. \label{fig:BestFit0and168}}
\end{figure}

In order to study the effects of injected methane upon the prompt scintillation light yield, we follow a very similar procedure to the one described in \cite{Jones:2013xy}.  We collect the areas of 40,000 pulses using a falling-edge trigger applied to PMT-UV.  The pulse area is evaluated and histogrammed online by the oscilloscope in a window which extends from -50ns to +100ns relative to the trigger time.  The histogram is then saved from the oscilloscope for further analysis offline.  We subtract the measured DC baseline and then fit the measured pulse area distribution with a function describing a poisson-like peak with corrections for source shadowing added to a falling power-law background.  Since the geometry of the alpha source holder is unchanged from that used in \cite{Jones:2013xy}, we use the same experimentally determined shadowing function from that study, more details of which can be found in \cite{Jones:2013xy}, section 3.  The pulse area spectrum and best fit function for two concentration points from this study are shown in figure \ref{fig:BestFit0and168}.  The light yield as a function of methane concentration is determined by the poisson mean of the alpha-induced peak at each concentration point.

\begin{figure}[t]
\begin{centering}
\includegraphics[width=0.8\columnwidth]{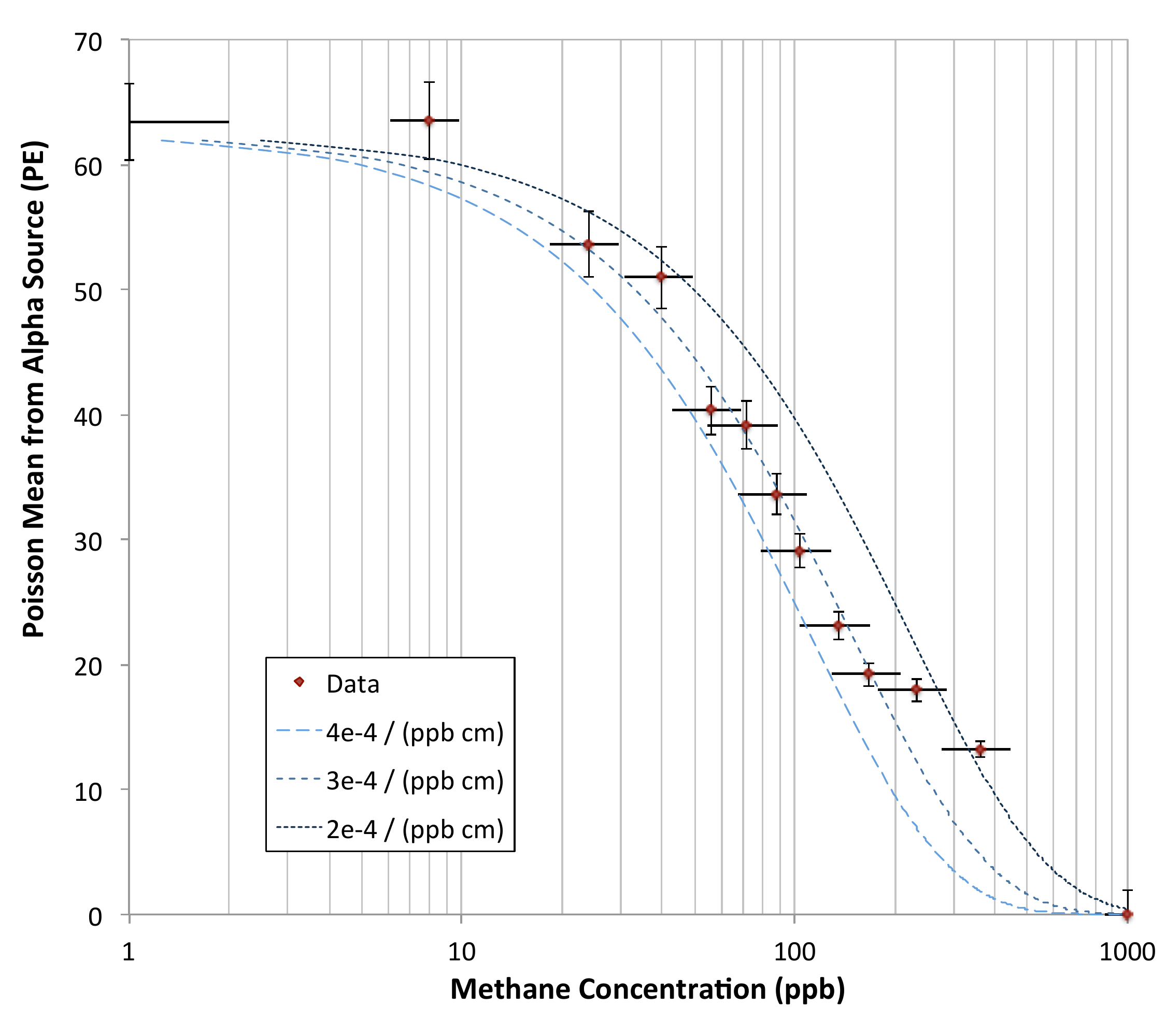}
\par\end{centering}

\caption{The measured prompt scintillation light yield from a monoenergetic polonium 210 alpha source as a function of methane concentration.  Horizontal error bars are given by the precision of the injection pressure gauge.  Vertical error bars are given by the single photoelectron scale added in quadrature with the peak fit error.  The left-most point is for pure argon.  The overlaid curves show the predicted attenuation behavior based on a ray tracing simulation for three characteristic attenuation strengths.  At the right-most point, the alpha-induced peak was no longer visible. \label{fig:PromptLightVsMethane}}
\end{figure}

\begin{figure}[t]
\begin{centering}
\includegraphics[width=1.0\columnwidth]{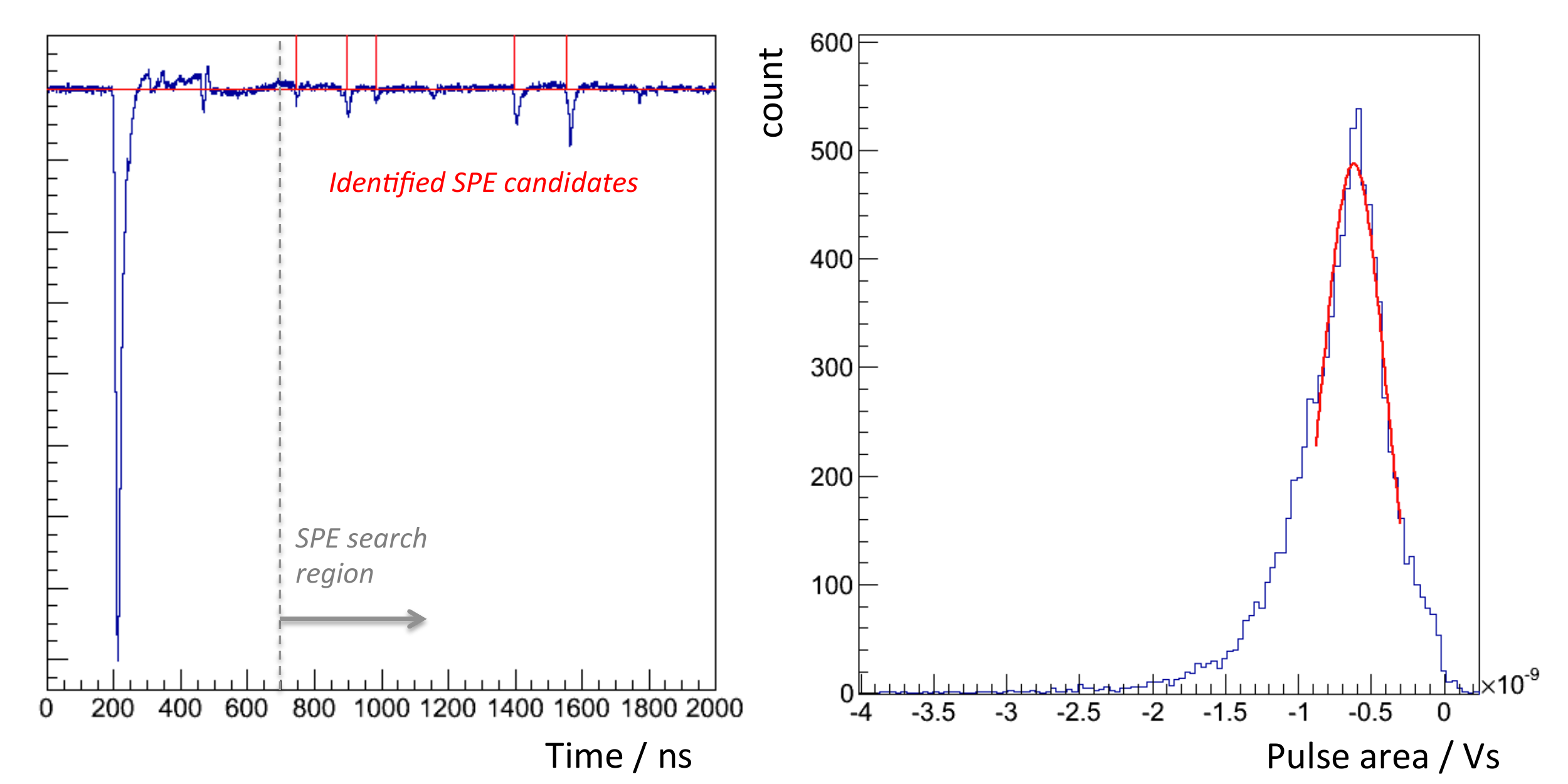}
\par\end{centering}

\caption{Left: a sample waveform showing where SPE candidates were found in the late scintillation light signal. Right: the distribution of SPE areas, with peak fit shown in red. \label{fig:SPEMethod}}
\end{figure}

The single photoelectron (SPE) scale which sets the pulse area normalization is measured at several points throughout the run using late scintillation light. To obtain the single photoelectron area distribution, 10,000 individual waveforms are analyzed using a peak-finding algorithm which isolates likely SPE pulses between 500 ns and 1400 ns following a falling-edge trigger, by identifying times where the waveform has a large and negative dV/dt over 5 samples.  An example waveform with the SPE candidate positions highlighted is shown in figure \ref{fig:SPEMethod}, left.  The baseline-subtracted areas of the SPE candidate pulses are evaluated by integrating from the beginning of the pulse until the amplitude falls back above threshold. The SPE areas are histogrammed, and a gaussian fit to the peak of this histogram gives the measured SPE scale.  A sample pulse area histogram with fitted peak is shown in figure \ref{fig:SPEMethod}, right.  Using this method, the measured SPE scale is found to be stable to within 2.5\% over the course of the run.

An alternative method of evaluating the SPE scale is to calculate the average area of all SPE candidate pulses found.  This method gives a SPE scale which is approximately 20\% higher than that obtained by fitting the peak, and this discrepancy is likely caused by contamination of the sample with pulses of which have contributions from more than one photoelectron, which appear in the large area tail of the distribution.  The SPE scale stability obtained with this method is 1.02\%, over the duration of the run.  In this analysis we use the peak-fitting method rather than the average area method to evaluate the SPE scale, as it is less sensitive to the effects of the multiple photoelectron pulse contamination.  Ultimately, our conclusions will rely only upon the stability of the SPE scale rather than its absolute value, and we include the measured SPE stability of 2.5\% as a systematic error on the normalization of each measured light yield.

In our previous publication \cite{Jones:2013xy} we also measured the SPE scale using a pulsed LED.  This method was found to give a SPE scale consistent with the peak-finding method to within 7\%.  The effects of shifts in the SPE scale by up to 20\% were investigated, and found to have no significant biasing effect on the relative prompt light yield measurement as a function of concentration.

The measured light yield as a function of methane concentration is shown in Figure \ref{fig:PromptLightVsMethane}.  The dominant systematic error on the light yield measurement is the SPE scale stability.  We see no observable light loss below 10 ppb, and then steady losses of scintillation light between 10 ppb and 1 ppm.  Beyond 1ppm of methane, the alpha peak was no longer observable.  Overlaid on the figure are the predictions of the expected light yield as a function of methane concentration for absorption strengths of 0.02\% , 0.03\% and 0.04\% $\mathrm{ppb^{-1} cm^{-1}}$ as determined by a ray tracing simulation of our apparatus, normalized to the emitted light intensity for clean argon.  These correspond to molecular absorption cross sections of $9.5\times10^{-18} ~\mathrm{cm}^{2}$, $1.4\times10^{-17} ~\mathrm{cm}^{2}$ and $1.9\times10^{-17} ~\mathrm{cm}^{2}$ respectively.  Our data are in approximate agreement with previous measurements of the methane absorption cross section at 128 nm, made at room temperature and pressure with pure methane gas \cite{WorldUVData}.  We address in more detail whether the losses we see can be attributed to absorption effects alone in section \ref{sec:QuenchingOrAbs}.

\begin{figure}[tb]
\begin{centering}
\includegraphics[width=0.6\columnwidth]{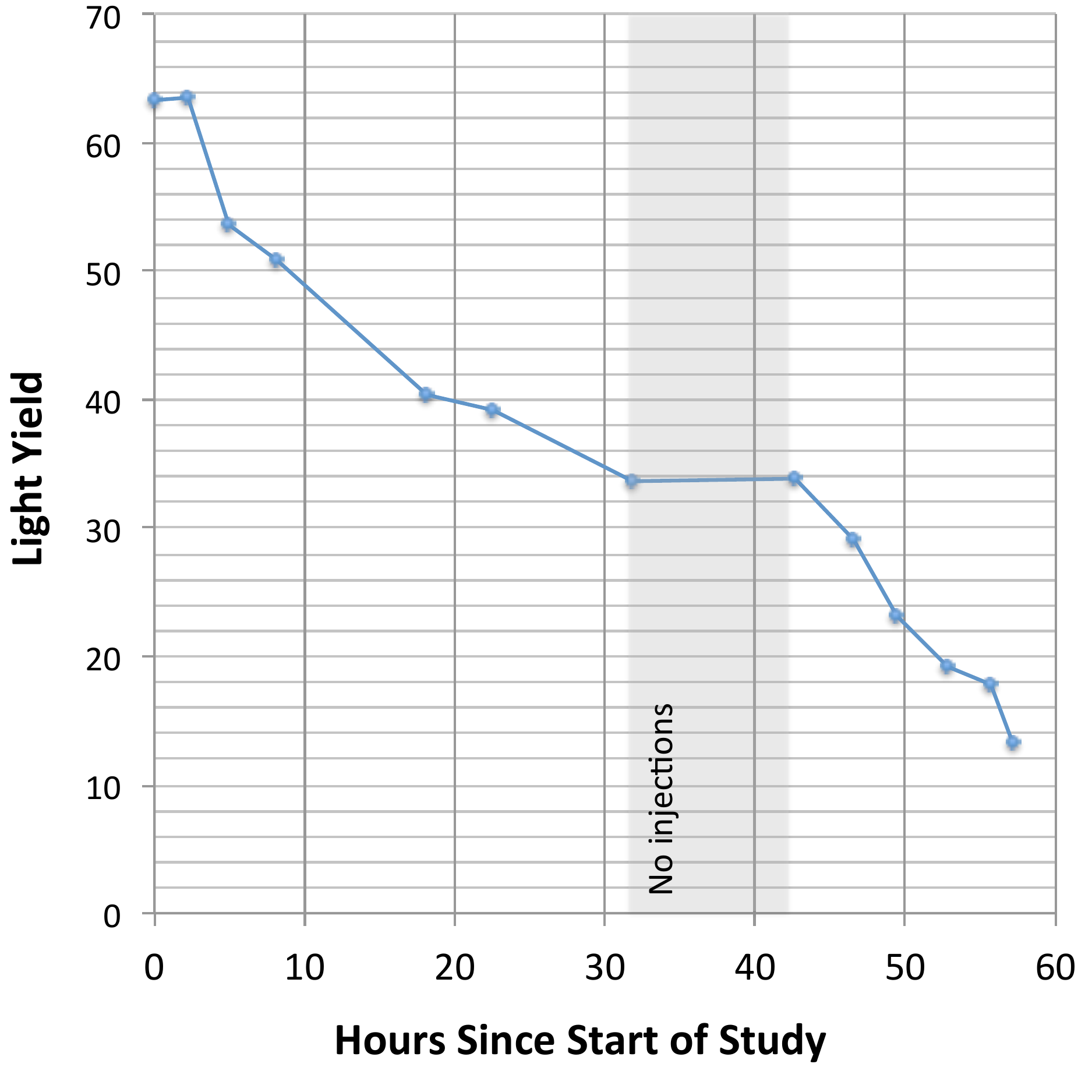}
\par\end{centering}

\caption{The prompt light yield as a function of time during this study.  For the period where no methane was injected, the light yield was found to be stable to within 1.2\% \label{fig:NoInjStability}}
\end{figure}

As a cross-check that the light losses observed were a result of methane injection and not outgassing of water or some other system instability, we collected similar datasets without injecting methane 12 hours apart during the middle of the run, and found stability of the light yield to within 1.2\%.  This is to be contrasted with the tens-of-percent light losses observed over 12 hour periods during methane injections.  These data are shown in Figure \ref{fig:NoInjStability}.  In a separate run, a single methane injection of 1ppm was made and the alpha-induced poisson-like peak was immediately extinguished.

\section{Argon Scintillation Time Constants, Quenching and Absorption \label{sec:QuenchingOrAbs}}

The light losses described in section \ref{sec:PromptYieldSection} could be the result of quenching, absorption or both.  In this context, quenching refers to a process where the excimers involved in the liquid argon scintillation process are dissociated by interaction with an impurity molecule without emitting a scintillation photon.  This mechanism affects late, triplet-state scintillation light much more strongly than prompt, singlet-state light, leading to a reduced time constant and late-to-total light ratio.  Absorption, on the other hand, refers to the loss of photons due to interactions with impurities during propagation between source and detection points, and affects both prompt and slow scintillation light equivalently.

To distinguish between these two types of processes we investigate the effects of methane concentration upon the time constant of the late liquid argon scintillation light.   If the large methane-induced losses observed in the prompt scintillation light are a result of a quenching process, we would expect large changes in the slow scintillation time constant at concentrations at or below those where the prompt peak is affected.  A pure absorption process would affect both components equally, leading to an overall reduction in light yield with no change of time constant.

\begin{figure}[tb]
\begin{centering}
\includegraphics[width=0.9\columnwidth]{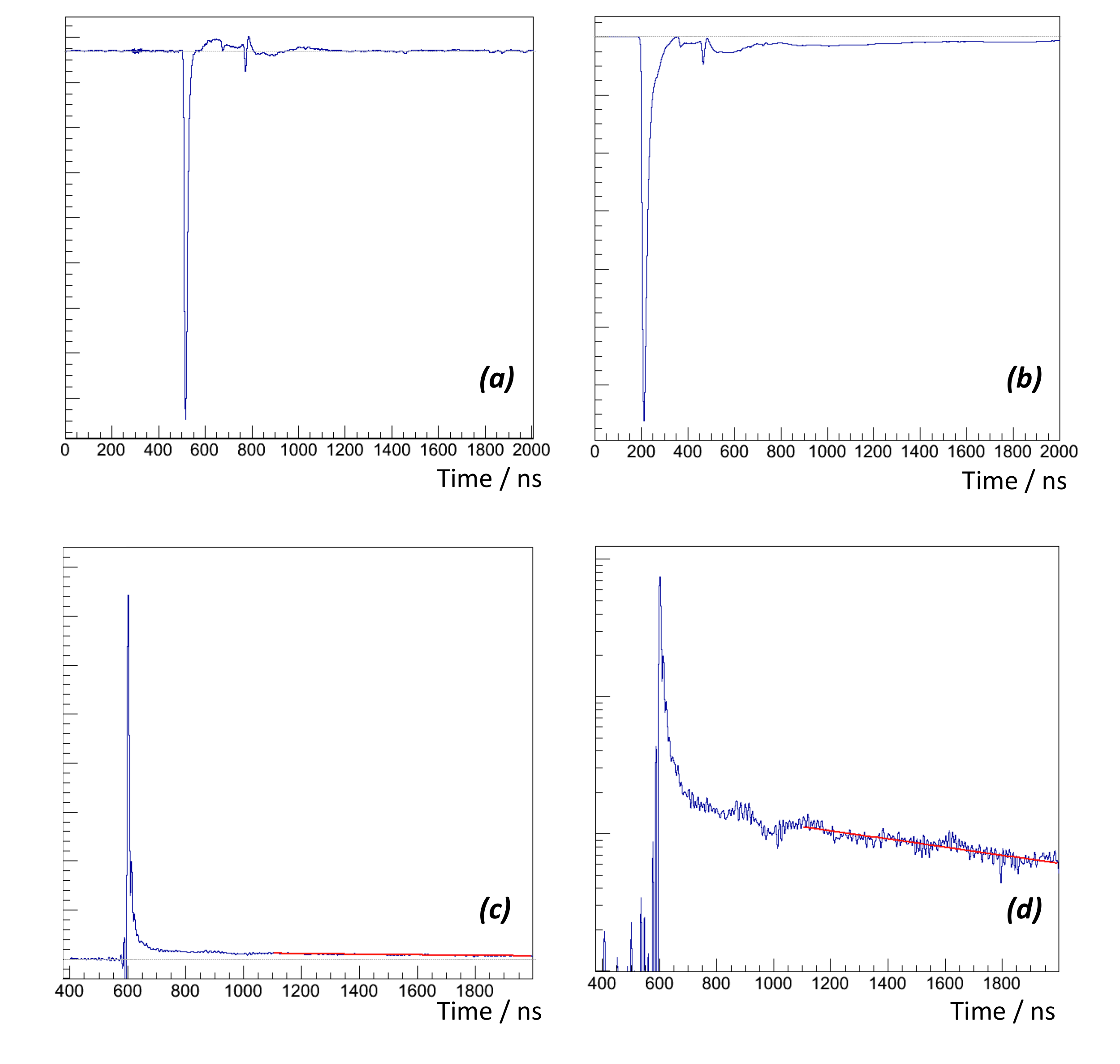}
\par\end{centering}

\caption{a) The averaged single photoelectron pulse shape, obtained with a pulsed LED.  b) The average raw pulse shape for scintillation deposits in clean argon.  c) The average deconvolved pulse shape for scintillation deposits in clean argon on a linear scale, with exponential fit overlaid in red. d) The average deconvolved pulse shape for scintillation deposits in clean argon on a logarithmic scale, with exponential fit overlaid in red.\label{fig:DeconvShapes}}
\end{figure}

The method used to extract the late scintillation time constant resembles the method reported in \cite{Acciarri:2008kv}.  The single photoelectron response shape was first measured using a pulsed LED, optically coupled to a fiber which points at the PMT-UV photocathode.  Then, 10,000 self-triggered and baseline-subtracted waveforms were summed to produce an average PMT-UV pulse shape for scintillation deposits at each concentration point.  The summed waveform undergoes some pre-processing to prevent ringing artifacts from deconvolution, including the application of a smooth sigmoid window function at the sample edges, and pre- and post-padding with zeroes to prevent wraparound effects.  The summed waveform is then deconvolved using the single photoelectron pulse shape as a deconvolution kernel, and an optimized Weiner filter combined with low-pass filter is applied to suppress noise and deconvolution artifacts.

\begin{figure}[tb]
\begin{centering}
\includegraphics[width=1.0\columnwidth]{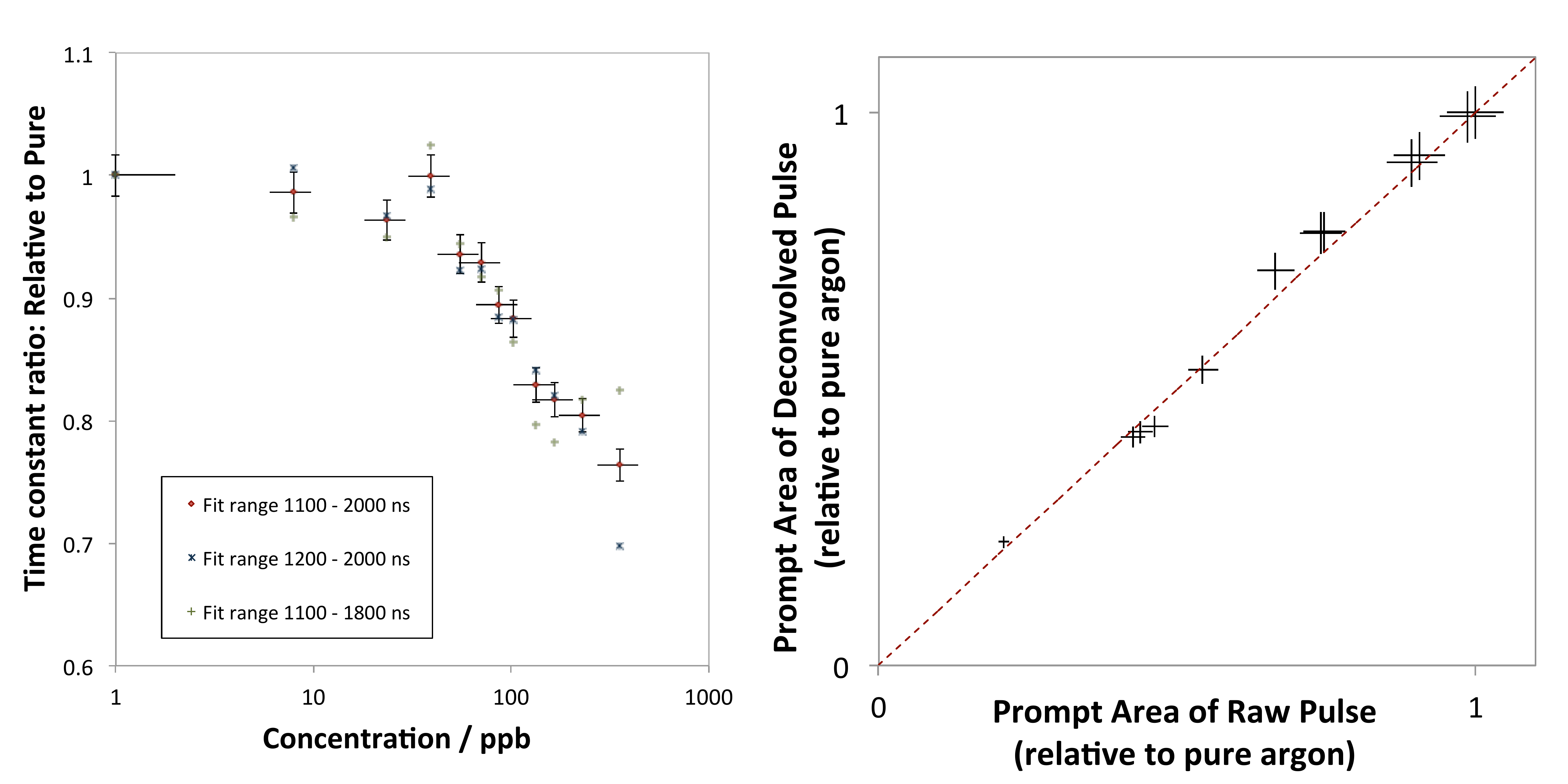}
\par\end{centering}

\caption{Left: Ratios of the time constant for contaminated argon to the time constant for pure argon, obtained using the full fit window (1100 - 2000 ns) and two reduced fit windows (1200 - 2000 ns and 1100 - 1800 ns) which yield pure argon time constants of 1462 $\pm$ 25 ns, 1627 $\pm$ 28 ns and 1455 $\pm$ 25 ns respectively. Right: Ratio of the average prompt areas of raw and deconvolved pulses at each concentration. \label{fig:RelativeTimeConsts}}
\end{figure}

This process removes all linear shape effects due to the PMT anode response, PMT undershoot, ringing in the PMT base electronics, and reflections in cables. The single photoelectron shape and average raw pulse shapes before and after deconvolution for clean argon are shown in figure \ref{fig:DeconvShapes}.  Nonlinear contributions to the pulse shape are not removed by this method.  As such, the reported late-light time constant obtained from this method is valid only to the extent that the PMT and electronics response is linear.  The level of system nonlinearity is very difficult to quantify.  For the purposes of this study, we assume that the system response is linear and thus the measured time constants are faithful.  Since our primary focus is upon changes in the late scintillation time constant rather than its absolute value, our main conclusions remain valid even in the presence of small nonlinear shape effects.  To ensure that pulse shape effects have not biased our prompt light measurements in section \ref{sec:PromptYieldSection} we can make a simple consistency check, comparing the average prompt areas obtained from raw and deconvolved pulses at each concentration point.  These data are shown in fig \ref{fig:RelativeTimeConsts}, right, and show good linearity within experimental uncertainties.

\begin{figure}[tb]
\begin{centering}
\includegraphics[width=1.0\columnwidth]{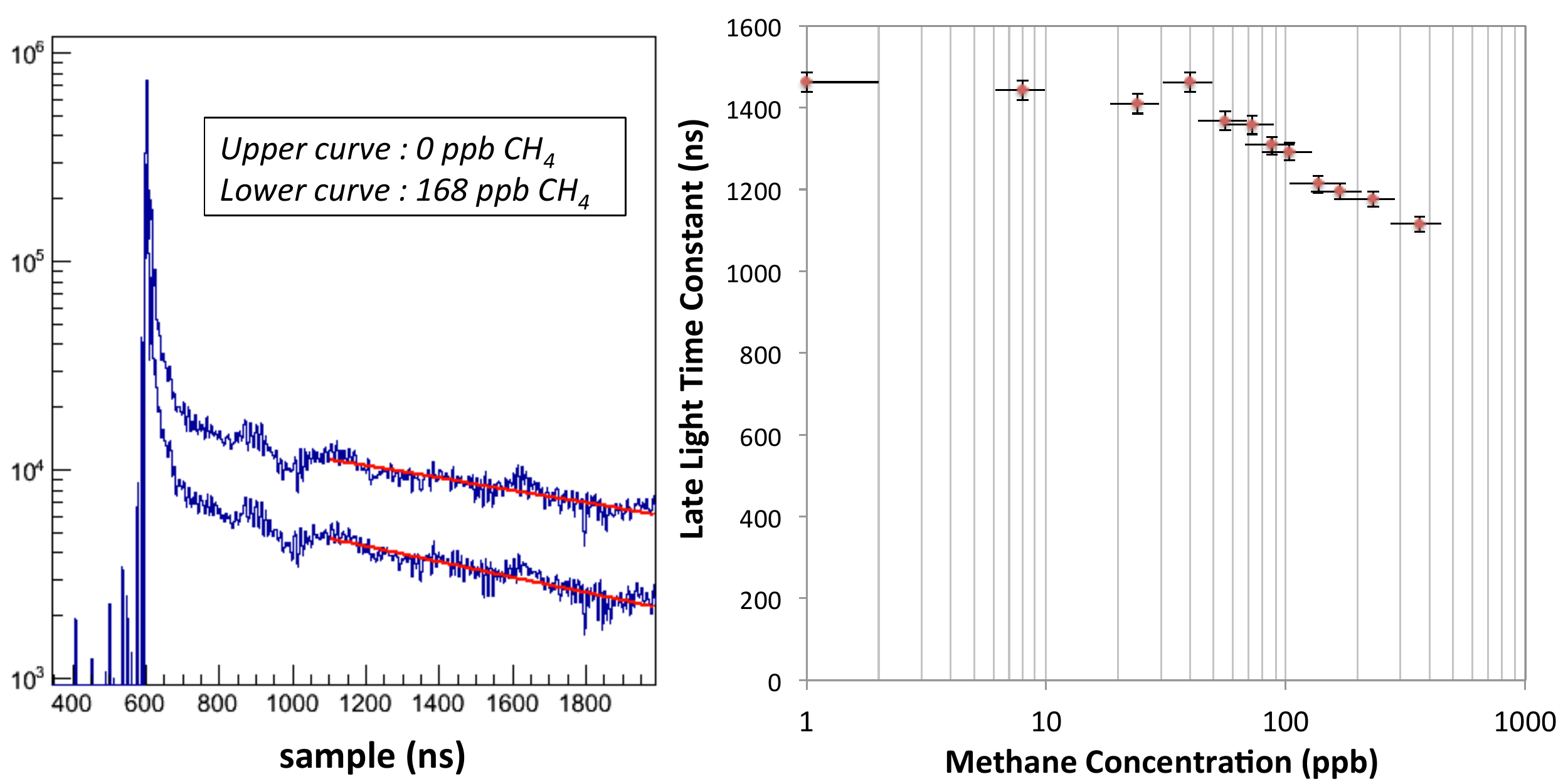}
\par\end{centering}

\caption{Left : Average deconvolved pulse shapes and exponential fits for two concentration points.  The red lines show the best-fit exponentials for each point.  Right: The measured late light time constant vs methane concentration.  Vertical error bars are given by the RMS discrepancy between fitting the first and second half of each dataset. The left-most point is for clean argon. \label{fig:TimeConstantVsMethane}}
\end{figure}

To extract the late-light time constant we fit the deconvolved pulse shape from 500~ns to 1400~ns after the prompt pulse with a falling exponential.  The 500~ns delay is chosen to avoid the effects of afterpulsing,  deconvolution artifacts from the sharp edge of the prompt peak and the effects of any possible intermediate scintillation time constants. Two deconvolved average waveforms and their corresponding fitted exponential functions are shown in Figure \ref{fig:TimeConstantVsMethane}, left.  The dependence of the measured time constant on methane concentration is shown in Figure \ref{fig:TimeConstantVsMethane}, right.

As has been discussed in appendix C of \cite{Acciarri:2008kv}, different definitions of the argon scintillation time constants and different fit methods used to extract them from deconvolved pulses can lead to different numerical values.  By fitting exponentials to shorter sections of the deconvolved pulse we were able to obtain time constants between 1450 and 1630 ns for clean argon.  However, the trend observed in the time constants as a function of methane concentration, quantified by comparing the time constant for pure argon to the time constant for contaminated argon, is found to be unchanged within experimental uncertainty regardless of which fit window is used.  Some examples are shown in figure \ref{fig:RelativeTimeConsts}, left.  In this study we are interested in this variation, rather than the absolute value of the time constant.  For this reason, the error bars shown in Figure \ref{fig:TimeConstantVsMethane} do not include this correlated systematic uncertainty, and are determined instead by the discrepancy between the time constants obtained from independently fitting the first and second half of the dataset for each concentration point, with the fit performed over the full fit region.

These data show very clearly that the majority of the light losses observed in the prompt scintillation light yield are not the result of a quenching process.  The late scintillation time constant remains long even when significant prompt light losses are observed.  Some evidence of a quenching process is observed at higher concentration values, but this effect is much too small to explain the observed light losses in the prompt peak.  The two curves of Figure \ref{fig:TimeConstantVsMethane} are for an unquenched and a heavily quenched point from this study.  The change in time constant between the 0 and 168 ppb concentration points is relatively small, whereas a significant loss of light yield for both prompt and late light components is clearly visible.  

We do not have the ability to test this apparent quenching behavior at much higher concentrations of methane in this detector, since the absorption effect is dominant and blocks a large fraction of the light required to observe the quenching.  A detailed study of this quenching process would be ideally performed in a much smaller detector, where light losses due to absorption between source and detector are smaller.  However, for any realistic scale neutrino or dark matter detector, the absorption effect is likely to be the dominant mechanism of light loss.

\section{Searching for Visible Re-emission Features}

It has been reported that a visible scintillation emission feature has been observed in gas-phase mixed argon and methane drift chambers \cite{Siegmund:1981un}.  The feature is suggested to be a 431~nm emission line resulting from interactions between argon excimers and various energy levels of methane molecules \cite{4336365}.  A competing non-radiative decay route at high methane densities is suggested to quench the emission, which is likely to suppress any such feature in a liquid phase mixture, where the molecular mean-free-path is much smaller than in the gas phase.  However, if some visible emission feature is present despite the strong absorption of 128~nm scintillation light, it could help to circumvent the problems associated with the absorption by allowing visible light to propagate across the detector, unabsorbed by methane. 

We search for visible emission features by measuring the areas of self-triggered PMT signals from PMT-Vis within a narrow window of a falling-edge trigger.  This PMT is optically isolated from most of the argon volume, but has a polonium disc source held 1 cm from the PMT face.  In pure argon, no features larger than the single photoelectron background peak are observed.

\begin{figure}[tb]
\begin{centering}
\includegraphics[width=0.8\columnwidth]{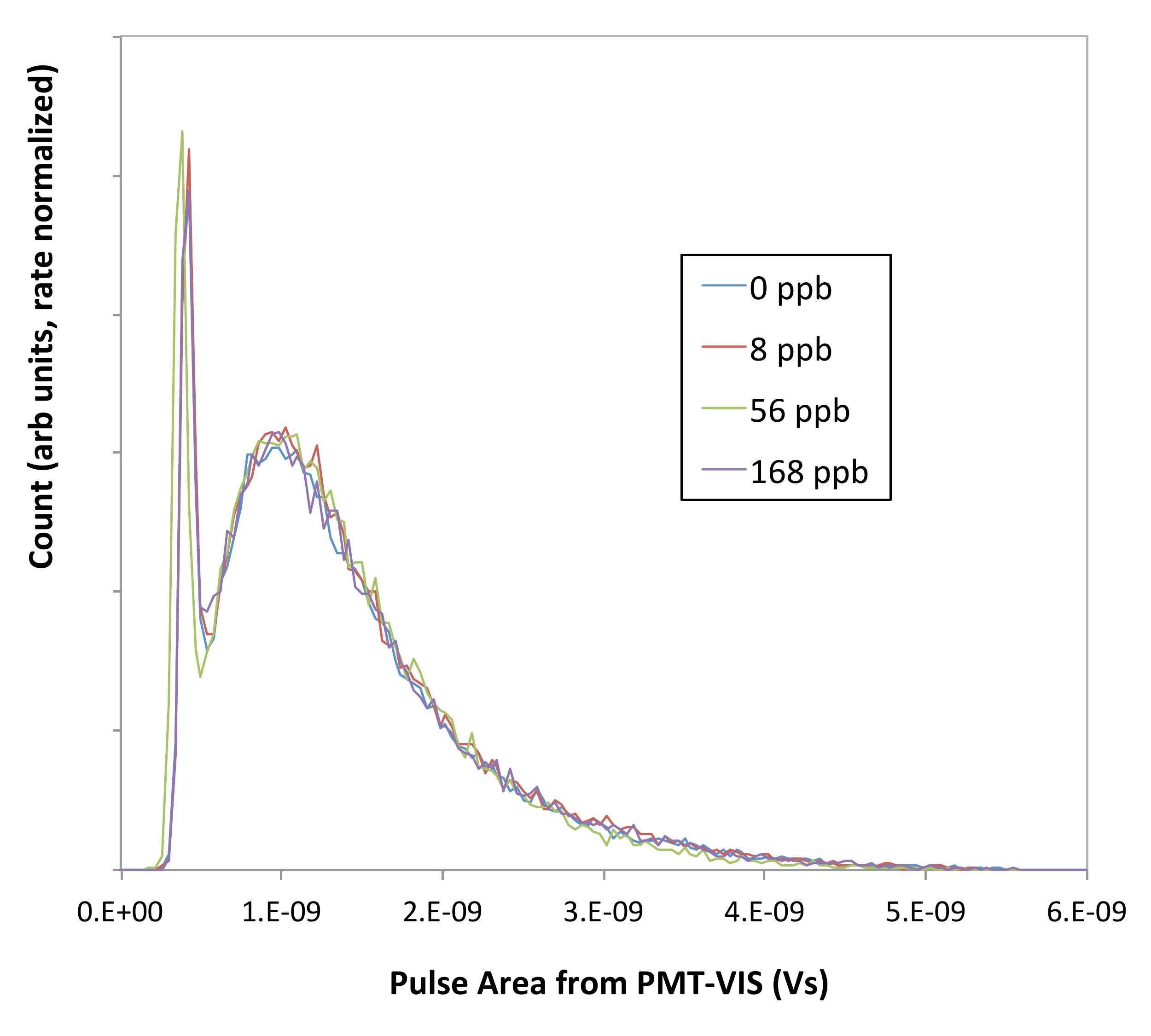}
\par\end{centering}

\caption{The rate-normalized charge per pulse measured on PMT-Vis, showing pedestal and SPE.  No SPE rate increase or higher intensity visible emission feature is observed as methane is injected. \label{fig:PMTVisLight}}
\end{figure}

The single photoelectron dark rate in clean argon is comparable to the alpha source rate; they are around 200 Hz and 100 Hz respectively.  This means that even triggering at the single photoelectron level, any alpha induced visible emission feature would be observable over background.  Rate-normalized histograms of pulse areas for several methane concentration points are shown in Figure \ref{fig:PMTVisLight}.  No change in the single photoelectron rate is observed, and no higher intensity features are detected for any concentration point.  These data were taken up to 168 ppb of methane, by which point 70\% of the 128~nm scintillation light was being absorbed.  We also performed a logarithmic scan of concentration points between 1~ppm and 0.1\% methane by mole, none of which have any observable 128~nm scintillation light, and saw no evidence of strong visible re-emission features at any concentration.  

\section{Implications for Neutrino and Dark Matter Experiments}

Our results show that significant light loss is to be expected when as little as tens of ppb of methane are mixed into liquid argon, due to ultraviolet absorption of 128~nm photons.  The loss of light is accompanied by no detectable visible re-emission, and as such, this poses a problem for any scintillation detectors operating in this methane concentration range.  

In the case of dark matter detectors using UAr, to eliminate the adverse affect of methane on argon scintillation, it is clear that the argon must contain less than 10 ppb of methane.  To reach this goal, hot getters can be used \cite{PrivateCommGetters}.  Since the light losses are attributed to an absorption process, the exact purity specification will depend on the geometry of the detector in question, with larger detectors requiring higher purity.

For TPC detectors aiming to use argon/methane mixtures in order to provide free protons, the detection of scintillation light is likely to be impracticable.  In order to provide a reasonable number of free protons for their physics goals, these detectors would need to contain methane at the few percent level or higher.  Whilst this type of argon/methane mixture has been shown to have acceptable properties for electron drift and charge collection \cite{Aprile1987273}, our studies show that such a detector would be blind to scintillation light, because of the methane absorption effect.

At higher concentrations, we have also seen evidence of an apparent quenching effect, which is inferred from a reduction of the late scintillation light time constant.  The configuration of our apparatus described in this paper is not well suited to studying the quenching effect in detail, since by the time the  effect is large, most of the emitted light has been absorbed between source and detector.  This is likely also to be true of any realstic neutrino or dark matter experiment, and so the practical implications of this quenching effect are limited.  A detailed study of the effect could be made in a test with a much smaller source to PMT path length, and with a data acquisition system capable of recording longer waveforms in order to capture many lifetimes of late scintillation light.

\acknowledgments

We would like to thank Clementine Jones for proofreading 
this paper, and Bill Miner, Ron Davis and the other technicians who have assisted us 
at the Proton Assembly Building,
Fermilab for their tireless hard work to provide us with cryogenic
facilities of the very highest standard. The authors thank the National
Science Foundation (NSF-PHY-1205175, NSF PHY-1211308 and NSF PHY-1242585) and Department Of Energy (DE-FG02-91ER40661).
This work was supported by the Fermi National Accelerator Laboratory,
which is operated by the Fermi Research Alliance, LLC under Contract
No. De-AC02- 07CH11359 with the United States Department of Energy.

\bibliographystyle{JHEP}
\bibliography{MethanePaper}{}

\end{document}